\documentclass[twocolumn,prl,showpacs,preprintnumbers,amsmath,amssymb]{revtex4}
\usepackage{graphicx}

\begin{document}

\title{Interstitial gas and density-segregation in vertically-vibrated granular media}


\author{M.\ Klein$^{1,2}$}
\author{L.L.\ Tsai$^{1,3}$}
\author{M.S.\ Rosen$^{1,2}$}
\author{D.\ Candela$^{4}$}
\author{T.\ Pavlin$^{1}$}
\altaffiliation[Currently at ]{Department of Circulation and Medical Imaging, Norwegian University of Science and Technology, Trondheim N-7489, Norway}
\author{R.L.\ Walsworth}
\affiliation{$^{1}$Harvard-Smithsonian Center for Astrophysics, Cambridge, Massachusetts 02138}
\affiliation{$^{2}$Department of Physics, Harvard University, Cambridge, Massachusetts 02138}
\affiliation{$^{3}$Harvard-MIT Division of Health Sciences and Technology, Cambridge, Massachusetts 02139}
\affiliation{$^{4}$Physics Department, University of Massachusetts, Amherst, Massachusetts 01003}

\date{\today.}

\begin{abstract}
We report experimental studies of the effect of interstitial gas on mass-density-segregation in a vertically-vibrated mixture of equal-sized bronze and glass spheres. Sufficiently strong vibration in the presence of interstitial gas induces vertical segregation into sharply separated bronze and glass layers.  We find that the segregated steady state ({\it i.e.},\ bronze or glass layer on top) is a sensitive function of gas pressure and viscosity, as well as vibration frequency and amplitude.  In particular, we identify distinct regimes of behavior that characterize the change from bronze-on-top to glass-on-top steady-state.
\end{abstract}

\maketitle

Vibrated granular media exhibit characteristics both similar to and distinct from solids, liquids, or
gases, with important open questions about the applicability of hydrodynamic theories and the emergence of patterns and order \cite{behringerRMP,duran,ristow,behringerHorzVert,bronemuzzioFreq,horzSeg}.  Recent experiments have investigated the striking effects  of interstitial gas on vibrated granular media  \cite{behringerPRL,mobius,yan,naylor,anotherJaeger,rhodes,mobius04,mobius05}.  In particular, it was recently demonstrated\ \cite{burtally01,burtally03,biswas} that interstitial gas (in contrast to vacuum) is necessary for spatial segregation (vertical layering) by particle {\it mass density} in a binary granular medium of equal-sized particles that is vertically vibrated in a sealed cell.  In this paper we report an experimental study of such density-segregation as a function of interstitial gas pressure and viscosity, as well as vibration frequency and amplitude.

Our experimental setup was similar to that of \cite{burtally01,burtally03}.  The granular mixture consisted of equal-sized solid spheres of bronze (ACU Powder International) and soda lime glass (MO-SCI Corp.)\ with diameter $d = 98 \pm 8\  \mu$m, where  90\% of both types of particles fell within this range. The respective mass densities of the bronze and glass are $8.9$ g/cm$^{3}$ and $2.5$ g/cm$^{3}$.  We used a volume mixing ratio of glass:bronze = 3:1, except where noted below.  To reduce moisture, we heated each mixed granular sample for 3--5 minutes and used an electrically grounded brass pan to reduce static charge buildup. We used such freshly-prepared samples for each data-taking run, which we performed over times $\lesssim$ $3$ hours.

\begin{figure}
\includegraphics[width=2in]{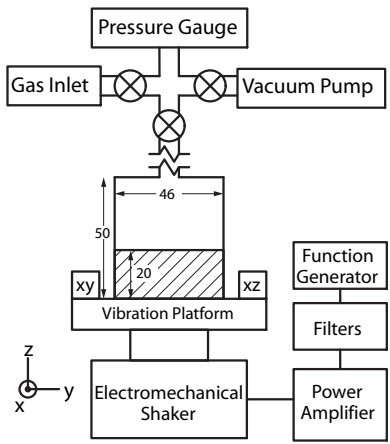}
\caption{Schematic
of the apparatus, not to scale.  Accelerometers are labeled by their sensitive axes ($xy$ and $xz$).  Dimensions are given in mm.  The cell depth along the $x$-axis is 10 mm (not shown)}
\label{fig1}
\end{figure}

We placed each sample inside a sealed, rectangular borosilicate glass cell (Spectrocell, Inc.)\ of width $46$ mm, height $50$ mm, and depth $10$ mm, mounted on a vibration platform (Fig.~\ref{fig1}).  The height of the granular mixture at rest in the sample cell was approximately $20$ mm for all data runs.  We used an electromechanical shaker (Labworks, Inc.\ ET-126) driven by a sinusoidal waveform to vibrate the cell vertically, nominally with vertical ($z$) position $z(t) = A \sin(2\pi ft)$ and peak velocity $v=2\pi Af=A\omega$.  We characterized this vibration by the frequency $f$ and dimensionless acceleration $\Gamma \equiv A\omega^{2}/g$, where $g$ is the acceleration of gravity.  We vibrated the cell over the range $f$ = 0--200 Hz and $\Gamma$ = 0--20.  Over this range the particles did not interact with the top of the sample cell.  A pair of accelerometers continuously monitored the vibration in three dimensions.  We ensured that the dimensionless horizontal accelerations ($\Gamma_{x}$ and $\Gamma_{y}$) were small ($< 0.2$, except as noted below), and limited variation in the dimensionless vertical acceleration ($\Gamma_{z}$) to no more than $0.1$ across the width of the platform.  A gas-delivery system controlled the interstitial gas type and pressure inside the sample cell ($P$ = 1--1000 Torr $\pm$ 3 Torr), without affecting the mechanical properties of the cell.  We performed experiments with either N$_{2}$ or Ne gas, which have respective viscosities $\eta$ = 17.9 and 32.1 $\mu$Pa$\cdot$s at room temperature (here $\eta$ is independent of gas pressure for $P \gtrsim 25$ Torr).

For our range of experimental conditions, the vibrated grains moved relative to the interstitial gas with Reynolds number $Re \approx (\rho d v / \eta) \approx$ 0.01--1, where $\rho$ is the gas mass density. Hence viscous drag was usually larger than inertial drag.  Importantly, there was no gas flow in and out of the sealed sample cell. Thus vertical gradients in the gas pressure were created across the vibrated particle bed inducing bulk gas flow through the granular medium (with gas flow velocity $v_g \sim \nabla P / \eta \sim P / \eta$ as given by application of Darcy's Law)  \cite{behringerPRL}.  To avoid hysteretic effects, we operated with the initial condition of a fully mixed, flat-topped sample, obtained by shaking under vacuum \cite{behringerPRL,burtally01,burtally03}.  We then admitted the desired gas; observed visually and with video recording and digital photography (20 $\mu$m resolution) the steady-state reached for long-term vibration with constant $\Gamma$ and $f$, as well as the behavior exhibited on the approach to the steady-state; removed the gas; and returned the sample to its initial mixed configuration before another run of the experiment.

\begin{figure}
\includegraphics[width=3in]{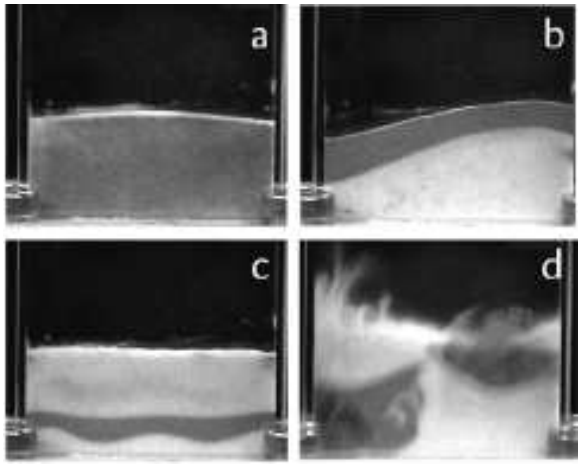}
\caption{Examples of steady-state behavior induced by vertical vibration.  (a) The mixed state, which occurs under vacuum and was used as the initial condition for most experiments.  (b) Bronze-on-top; N$_{2}$ interstitial gas at $P=500$ Torr, $\Gamma=8.7$, $f=90$ Hz. Digital images show the boundary between the bronze and glass layers to be only one or two grain diameters thick.  (c) Glass-on-top; same parameters as (b), except $\Gamma=10.8$.  Often a glass layer also forms below the bronze layer.  (d) Unstable pattern formation characteristic of very large $\Gamma$; N$_{2}$ interstitial gas at $P=800$ Torr, $\Gamma=14$,
$f=50$ Hz.}
\label{fig2}
\end{figure}

Using this apparatus, we observed steady-state patterns of vibration-induced density-segregation such as ``bronze-on-top'' and ``glass-on-top'' (see Fig.~\ref{fig2}) consistent with the schematic results described in\ \cite{burtally01,burtally03}. We then studied the dependence of density-segregation and particle mobility on gas properties. We found that the heavier bronze particles are more mobile than the glass spheres when vertically vibrated in the presence of interstitial gas, which is consistent with particle deceleration due to gas-drag scaling inversely with particle mass. For example, we observed that vibrated bronze particles leave the sample surface and reach a noticeably greater height (about 3 times higher) than glass particles near the surface. Also, we determined the local expansion (dilation) of the granular bed under vibration to be greater for regions with higher bronze concentration: images showed these regions to have a greater total volume than given by the sample's glass:bronze mixing ratio.  Consistent with previous experiment \cite{burtally03} and simulation \cite{biswas}, we found indirect evidence that the lighter glass particles are preferentially dragged along with bulk gas flow:  {\it e.g.}, the glass-on-top state {\it never occurred} when bulk gas flow due to gas pressure gradients was inhibited by placing the sample in a cell with a gas-permeable bottom \cite{porouscell}.  In addition, with this gas-permeable-bottom cell we found that overall particle mobility and bed expansion were diminished.  Correspondingly, $\Gamma$ had to be increased by about a factor of two (by increasing the  amplitude $A$) to induce bronze-on-top segregation. Further, as described in \cite{burtally03} all forms of density segregation --- including both bronze-on-top and glass-on-top --- were suppressed for a cell with a gas-permeable bottom {\it and} top. These results are consistent with bed expansion and particle mobility being enhanced by bulk gas flow through the sample.

\begin{figure}
\includegraphics[width=2.6in]{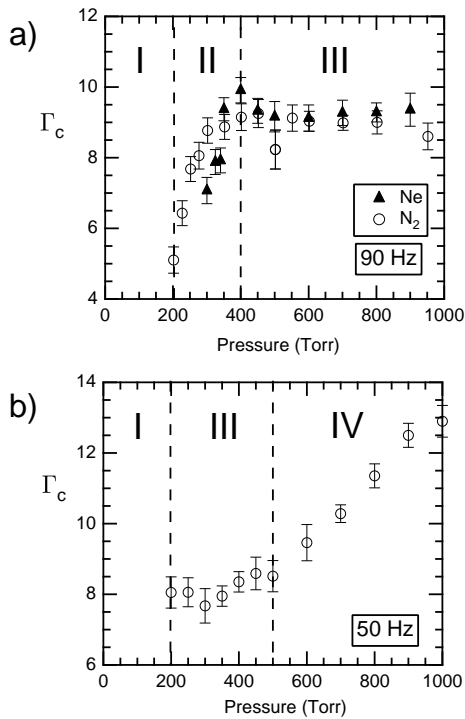}
\caption{Measured gas pressure dependence of the critical value of the dimensionless acceleration ($\Gamma_{c}$) delineating the change from the bronze-on-top to glass-on-top steady states.  (a) $f=90$ Hz vibration frequency, comparing N$_{2}$ and Ne interstitial gas.  (b) $f=50$ Hz, N$_{2}$ gas only; the glass-on-top state does not occur at this frequency for Ne.  Four approximate pressure regimes of characteristic $\Gamma_{c}$ behavior are indicated: ({\it I}) glass-on-top segregation does not occur at low gas pressure; ({\it II}) glass-on-top segregation occurs with the sample approximately flat on top; ({\it III}) glass-on-top segregation occurs via a pressure-independent spill-over mechanism; ({\it IV}) glass-on-top segregation occurs via pressure-dependent spill-over with inertial drag nominally comparable to viscous drag.}
\label{fig3}
\end{figure}

To characterize the competing effects of gas drag --- {\it i.e.},\ greater mobility of the heavier bronze particles and preferential dragging of the lighter glass particles with bulk gas flow --- we used the sealed sample cell and performed experiments focusing on the typically sharp boundary in vibration-amplitude parameter space between the stable bronze- and glass-on-top states.  In particular, we determined the critical value of the dimensionless acceleration ($\Gamma_{c}$) that delineates the change from bronze-on-top to glass-on-top steady states, as a function of gas pressure ($P$), vibration frequency (up to 200 Hz), and gas viscosity. Figure~\ref{fig3} shows examples of the measured dependence of $\Gamma_{c}$ on $P$ for typical high and low vibration frequencies ($90$ and $50$ Hz) and gases of differing viscosity (Ne and N$_{2}$).  As discussed below, we observed a dramatic difference in the dependence of segregation behavior on gas viscosity at $f=50$ Hz, but little difference at $f=90$ Hz.  Specifically, the glass-on-top state does not occur at $f=50$ Hz for Ne gas at any pressure.

Each value of $\Gamma_{c}$, plotted in Fig.~\ref{fig3}, is an average between the highest $\Gamma$ to yield bronze on top ($\Gamma_{b-hi}$) and the lowest $\Gamma$ to yield glass-on-top ($\Gamma_{g-lo}$), as determined by repeated sets of experimental runs.  Uncertainty in the determination of $\Gamma_{c}$ comes from several sources: irreproducibility of $\Gamma_{b-hi}$ and $\Gamma_{g-lo}$, which contributed an uncertainty $\delta\Gamma_{1}\approx 0.1$; spatial variation of $\Gamma_{b-hi}$ and $\Gamma_{g-lo}$ across the width of the platform, with associated uncertainty $\delta\Gamma_{2}\approx 0.1$; and the systematic offset $\delta\Gamma_{3}=\Gamma_{b-hi}-\Gamma_{g-lo}$, which varied from $\approx$ 0--0.5 for differing experimental conditions.  The total uncertainty for each plotted value of $\Gamma_{c}$ is $\Delta\Gamma_{c}=(\delta\Gamma_{1}+\delta\Gamma_{2})^{1/2}+\delta\Gamma_{3}$.

>From the measurements of $\Gamma_{c}$ vs.\ $P$ and associated observations of bed dynamics we identify four characteristic regimes of steady-state behavior, as indicated in Fig.~\ref{fig3} and summarized in the following.

{\it Regime I (no glass-on-top steady state at low gas pressure)} --- The glass-on-top steady state does not occur at low gas pressure (below about 200 Torr for N$_{2}$ and 300 Torr for Ne), so $\Gamma_{c}$ is undefined.  In this low pressure regime we observed minimal bed dilation and low mobility of the glass particles, which is consistent with relatively small drag from bulk gas flow driven through the particle bed by vibration-induced pressure gradients (since the gas flow velocity scales as $v_g \sim \nabla P \sim P$). Note that at very low gas pressure ($\lesssim 50$ Torr) we found that the bronze-on-top steady-state also does not occur.

\begin{figure}
\includegraphics[width=3.1in]{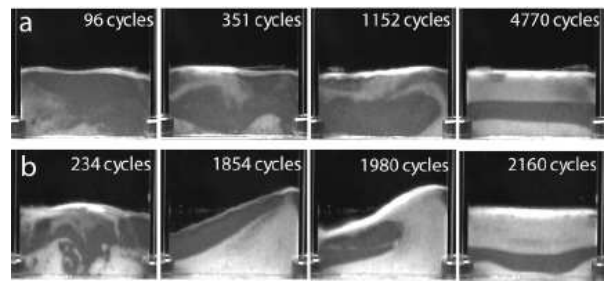}
\caption{Time-resolved images illustrating evolution from an initial mixed state [like Fig.~\ref{fig2}(a)] toward a glass-on-top steady-state.  (a) Regime II: N$_{2}$ gas, $P=250$ Torr, $\Gamma=8.9$, $f=90$ Hz.  (b) Regime III: N$_{2}$ gas, $P=500$ Torr, $\Gamma=10.8$, $f=90$ Hz.}
\label{fig4}
\end{figure}

{\it Regime II (glass-on-top occurs with minimal sample heaping)} --- At moderate gas pressures and high vibration frequencies such as $f=90$ Hz, both bronze-on-top and glass-on-top steady-states occur with the top of the vibrated particle bed remaining approximately flat; {\it i.e.},\ with minimal heaping, see Fig.~\ref{fig4}(a). As $P$ is increased, the observed bed dilation and particle mobility increase, the bronze-on-top steady-state is reached more quickly (for $\Gamma < \Gamma_{c}$), and $\Gamma_{c}$ becomes larger. Similar dynamics, steady states, and values for $\Gamma_{c}$ are found for N$_{2}$ and Ne interstitial gases. See Fig.~\ref{fig3}(a).  This observed pressure-dependence and approximate viscosity-independence for $\Gamma_{c}$ is consistent with viscous drag due to bulk gas flow through the particle bed playing a dominant role \cite{biswas}, which implies that the drag force on the particles scales as $F_{v} \sim \eta v_g \sim P$ since application of Darcy's Law gives $v_g \sim \nabla P / \eta \sim P / \eta$ in a sealed sample cell \cite{behringerPRL}. At lower vibration frequencies such as $f=50$ Hz, Regime II behavior does not occur, since for the relatively larger vibration and gas flow velocities ($\approx \Gamma g/\omega$) sample heaping already ensues in Regime I,  {\it i.e.}, at gas pressures below that required for glass-on-top segregation.

{\it Regime III (glass-on-top occurs via pressure-independent sample spill-over)} --- Above a gas pressure threshold, the glass-on-top state is reached via a ``spill-over'' mechanism in the presence of sample heaping, as illustrated in Fig.~\ref{fig4}(b).  For $\Gamma \ge \Gamma_{c}$ the sample quickly forms a transient bronze-on-top layer, as heaping occurs across the width of the bed.  The heaping builds until the bronze layer surpasses the angle of marginal stability \cite{ristow}. Particles slide down the heap, with partial replenishment via convective upflow in the bronze layer.  Glass particles then break through at the top of the heap and spill over, forming the
glass-on-top steady-state.  Particle-bed heaping is known to arise from vibration-induced pressure gradients and bulk gas flow, with a steepness that is largely pressure- and viscosity-independent for the pressures used here \cite{behringerPRL}.  Correspondingly, our measurements of $\Gamma_{c}$ show little dependence on $P$ and $\eta$ in Regime III at higher vibration frequencies such as $f$ = 90 Hz.  For lower vibration frequencies, however, we observed that the spill-over mechanism is inhibited relative to the convective replenishment of the bronze layer.  In particular, the glass-on-top state cannot be achieved for Ne gas and $f \lesssim 75$ Hz, nor for N$_{2}$ gas and $f \lesssim 40$ Hz, which is consistent with the opposite scaling of the viscous drag force with $\eta$ and $f$: $F_{v} \sim \eta v_g \sim \Gamma_{c}\eta / f$, with $\Gamma_{c}$ near the typical value for Regime III.

\begin{figure}[tbh]
\includegraphics[width=2.6in]{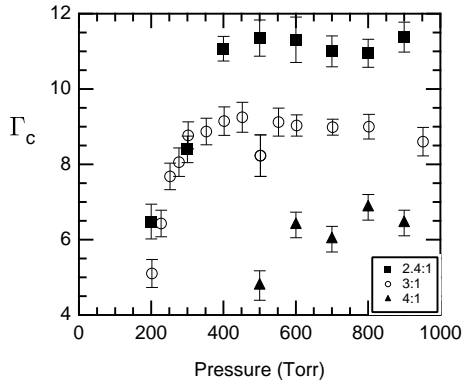}
\caption{Measured $\Gamma_{c}$ variation with N$_{2}$ gas pressure at $f=90$ Hz, for three glass:bronze mixing ratios, 2.4:1 (squares), 3:1 (circles), and 4:1 (triangles).   As the bronze fraction grows, the thickness of the bronze-on-top layer increases, which suppresses the mobility of the underlying glass particles and increases $\Gamma_{c}$.}
\label{fig5}
\end{figure}

{\it Regime IV (glass-on-top occurs with pressure-dependent sample spill-over)} --- At low vibration frequency and high gas pressure the observed difference in bronze and glass mobility grows with pressure, which inhibits the spill-over mechanism to the glass-on-top steady-state and yields an approximately linear dependence of $\Gamma_{c}$ on $P$. See Fig.~\ref{fig3}(b). This pressure dependence may result from inertial gas drag on the particles, which according to a simple estimate ($Re \sim P / \eta f \sim$ 1) becomes significant in this regime.

Further study will be needed to establish a full, quantitative understanding of the effect of interstitial gas on density-segregation in vertically-vibrated granular media. For example, density-segregation can depend in detail on system parameters such as the particle mixing ratio.  In our experiments we found that a bronze-on-top layer greatly suppresses the mobility of the underlying glass particles; and the thicker the bronze layer the greater the suppression of glass mobility.  One consequence of this bronze-on-top suppression is demonstrated in Fig.~\ref{fig5}: introducing more (less) bronze into the system requires a larger (smaller) $\Gamma_{c}$ to bring glass to the top, as the glass-on-top steady-state is generally preceded by a transient bronze-on-top condition.  In future work we plan to employ NMR/MRI techniques to map grain motion as well as gas pressure and flow dynamics, by using NMR-detectable grains \cite{NMRseeds} and hyperpolarized noble gas \cite{NMRlaserpolarized}, respectively.  We are also performing NMR investigations of fluidized beds \cite{WalsFluidBeds} and may extend these studies to segregation phenomena \cite{howleyglasser,watersegregation}.

This work was supported by NSF Grant No. CTS-031006.


\begin{thebibliography}{13}
\bibitem{porouscell}
The cell was made of Lexan with a bottom of steel mesh.  Sealing the bottom restored all segregation phenomena observed with the sealed glass cell.
\end{thebibliography}
\end{document}